\documentclass[a4paper,11pt]{article}
\usepackage{pos}

\usepackage{wrapfig}

\title{Tagging the initial state gluon in the Z+jet process}

\author*[a,b]{Simone Caletti}

\affiliation[a]{Dipartimento di Fisica, Università di Genova,\\
Via Dodecaneso 33, Genoa, Italy}
\affiliation[b]{INFN, sezione di Genova}

\emailAdd{simone.caletti@ge.infn.it}

\abstract{We develop a jet flavour tagger in the contest of electroweak boson production in association with jets. Here the jet with the highest transverse momentum is considered and a simple cut on a jet angularity may serve as a tagger for the flavour of the jet. This is an infrared and collinear (IRC) safe procedure thus it is well-defined from a theoretical viewpoint. Jet angularities exibit a property called Casimir Scaling (CS) \cite{Larkoski:2014pca} at the leading logarithmic (LL) accuracy. Here we consider also the first deviation from the typical CS behaviour because we use the most recent jet angularities calculations developed in \cite{Caletti:2021oor}. Tagging the leading jet as quark-initiated permits us to enhance the initial-state gluon purity. We will consider transverse momentum distributions for tagged and no-tagged jets, with and without grooming. They could be potentially interesting observables to probe gluonic degrees of freedom of the colliding protons. In particular it may be worth to investigate if such a study might be a new handle on the determination of the gluon parton distribution function (PDF).
}

\FullConference{%
 
 The Ninth Annual Conference on Large Hadron Collider Physics - LHCP2021\\
 7-12 June 2021\\
 Online
}

\begin{document}
\maketitle

\section{Quark/gluon discrimination and jet angularities}
Jets are robust tools that allow us to describe short-distance collisions involving quarks and gluons. A well-defined procedure to assign a flavour label to a jet would be beneficial, for instance, for several application in "beyond the Standard Model" (BSM) searches\footnote{See \cite{Gras:2017jty} references from $2$ to $19$ for more concrete examples.}. However quark/gluon tagging is an intrisically ill-defined problem because partons carry color while jets are composed of color-singlet hadrons, so the the flavour label is fundamentally ambiguous. As discussed in \cite{Gras:2017jty} one can indeed create a well-defined quark/gluon tagging procedure based on unambiguous hadron-level measurement. In short, what we mean by a quark/gluon jet is a phase space region (as defined by an unambiguous hadronic fiducial cross section measurement) that yields as enriched  sample of quarks/gluons (as interpreted by some suitable, though fundamentally ambiguous, criterion). Such a criterion is developed in this context performing a simple cut on the jet angularity distribution.
Jet angularities are observables which probe the energy flow within a jet. They are defined as
\begin{equation}
\lambda_\alpha^\kappa=\sum_{j\in\text{jet}}\left(\frac{p_{T,j}}{\sum_{j\in\text{jet}}p_{T,j}}\right)^\kappa\left(\frac{\Delta_j}{R}\right)^\alpha
\end{equation}
where $\kappa$ is fixed to be $1$ and $\alpha>0$ because of infrared and collinear (IRC) safety. The procedure to select the $\lambda_\text{cut}$ value will be discussed in Section 3.

\begin{figure}[h!]
\centering
\begin{minipage}{.45\linewidth}
  \includegraphics[width=\linewidth]{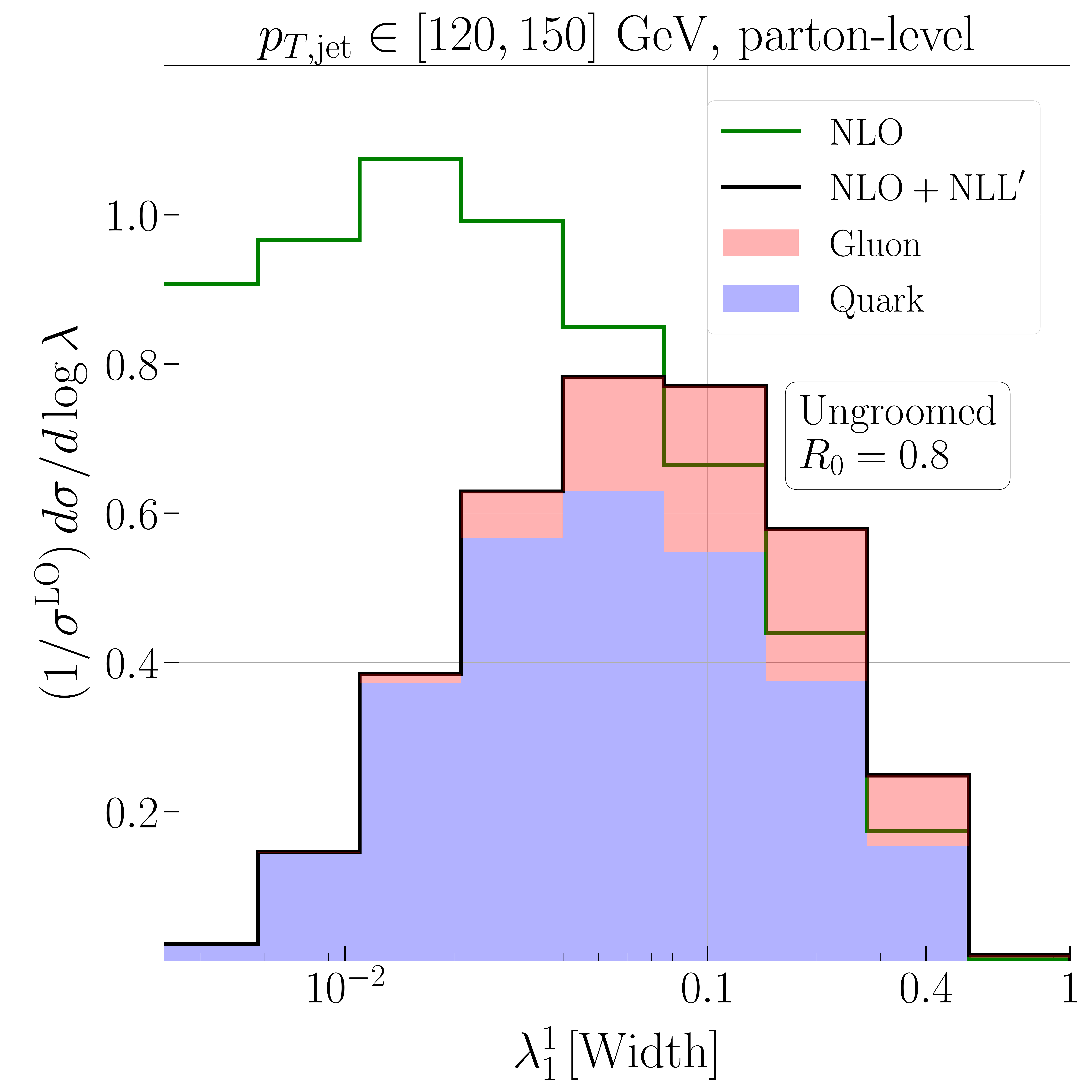}
  \captionof{figure}{NLO and NLO+NLL' predictions for the Width distribution, i.e. angularity with $\alpha=1$. From \cite{Caletti:2021oor}.}
  \label{fig:width}
\end{minipage}
\hspace{.05\linewidth}
\begin{minipage}{.45\linewidth}
  \includegraphics[width=\linewidth]{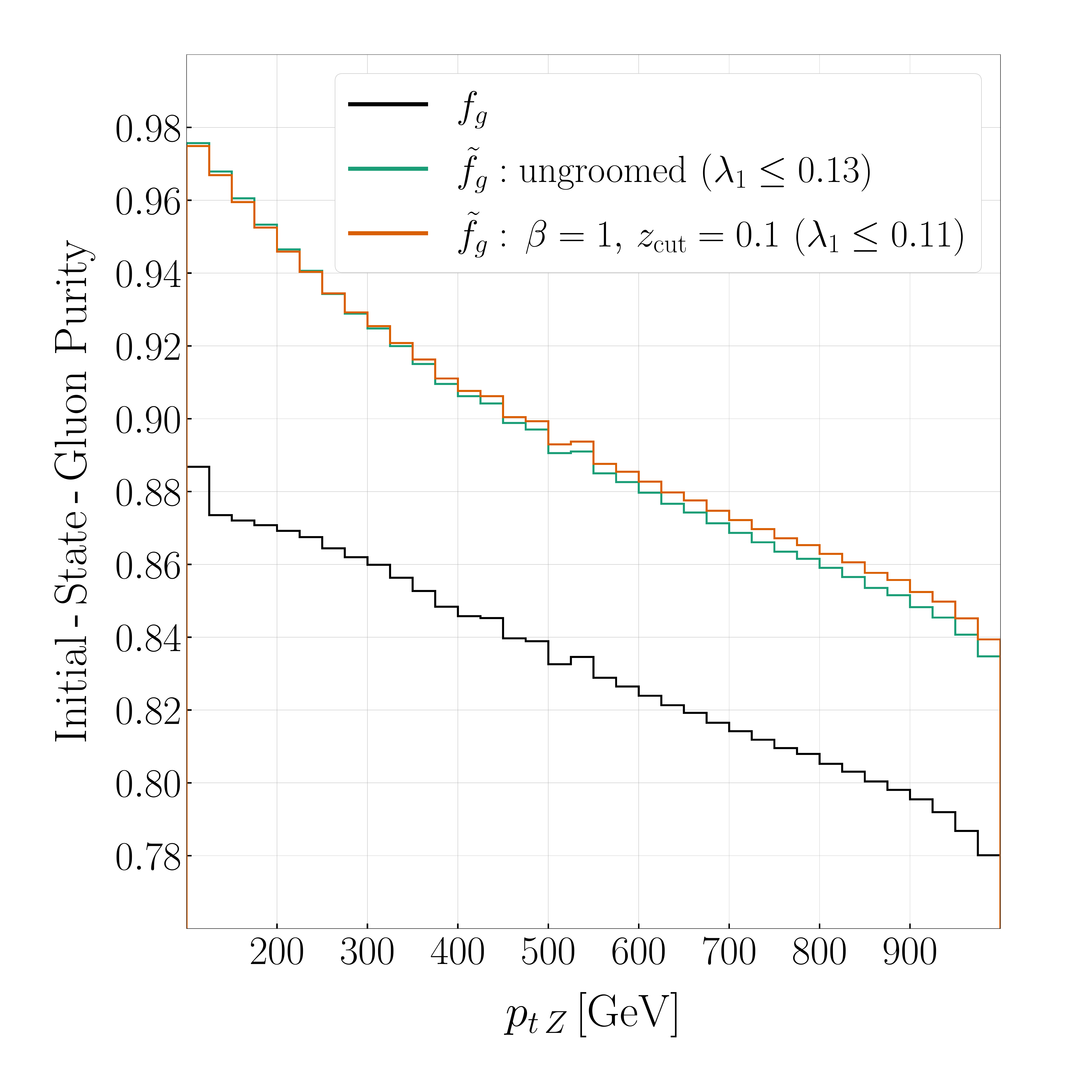}
  \captionof{figure}{Initial-state gluon purity before ($f_g$) and after ($\tilde{f_g}$) tagging obtained with PYTHIA simulations, as a function of the transverse momentum of the Z boson. From \cite{Caletti:2021ysv}.}
  \label{fig:purity}
\end{minipage}
\end{figure}

\section{Enriched parton samples in the Z+jet process}
Jet angularities have been recently calculated at NLO+NLL' accuracy\footnote{Next-to-leading order matched with next-to-leading logarithmic resummation. Matching to fixed-order is performed in such a way to ensure what is usually referred to as NLL' accuracy, with the prime sign. For further informations see \cite{Caletti:2021oor}.} in \cite{Caletti:2021oor}. The results for $\alpha=1$ (Width) is shown in Fig. \ref{fig:width}. Here the quark and the gluon contributions are separated using the BSZ flavour-$k_t$ algorithm \cite{Banfi:2006hf}. It is evident that the low-$\lambda_1^1$ tail is entirely dominated by quark jets, while a more significant contribution from gluon jets occurs for larger values of the angularity. This fact is still valid also after applying grooming. In particular, in \cite{Caletti:2021oor, Caletti:2021ysv} Soft Drop (SD) with $z_\text{cut}=0.1$ and different values of $\beta$ is considered. This is a confirmation that a cut on the jet angularity can serve as a theoretically well-defined and IRC-safe quark/gluon discriminant as pointed out, for instance, in \cite{Gras:2017jty, Larkoski:2014pca, Andersen:2016qtm, Amoroso:2020lgh}.\\

Now consider the Feynman diagrams which contribute to the LO partonic cross-section for the Z+jet production at the LHC. There are just two tree-level diagrams: 
\begin{itemize}
\item $qq \rightarrow gZ$;
\item $qg \rightarrow qZ$.
\end{itemize}
Now we define the \emph{initial-state gluon purity}
\begin{equation}
f_g=\frac{\sigma_{qg}}{\sigma_{qq}+\sigma_{qg}}\,,
\end{equation}
which represents, in this LO picture, the fraction of events having a gluon in the initial state.
We plot the initial-state gluon purity as a function of the transverse momentum of the Z boson using the PYTHIA Monte Carlo event generator. Notice that gluons contribute almost $80\%$ of the times to the plain process (without any tagging procedure) wihch is represented by the black line in Fig. \ref{fig:purity}. 
Since at the LO a quark-jet in the final state is $100\%$ correlated to the presence of a gluon in the initial state, we will use this relation between initial and final states at the parton level to pick initial-state gluons simply tagging the final-state jet as a quark-jet. In order to do that we have to define a tagging procedure accordingly to Section 1. This is the aim of the next Section.

\section{Cut selection and the enhanced purity}
\begin{figure}[]
\centering
\begin{minipage}{.45\linewidth}
  \includegraphics[width=\linewidth]{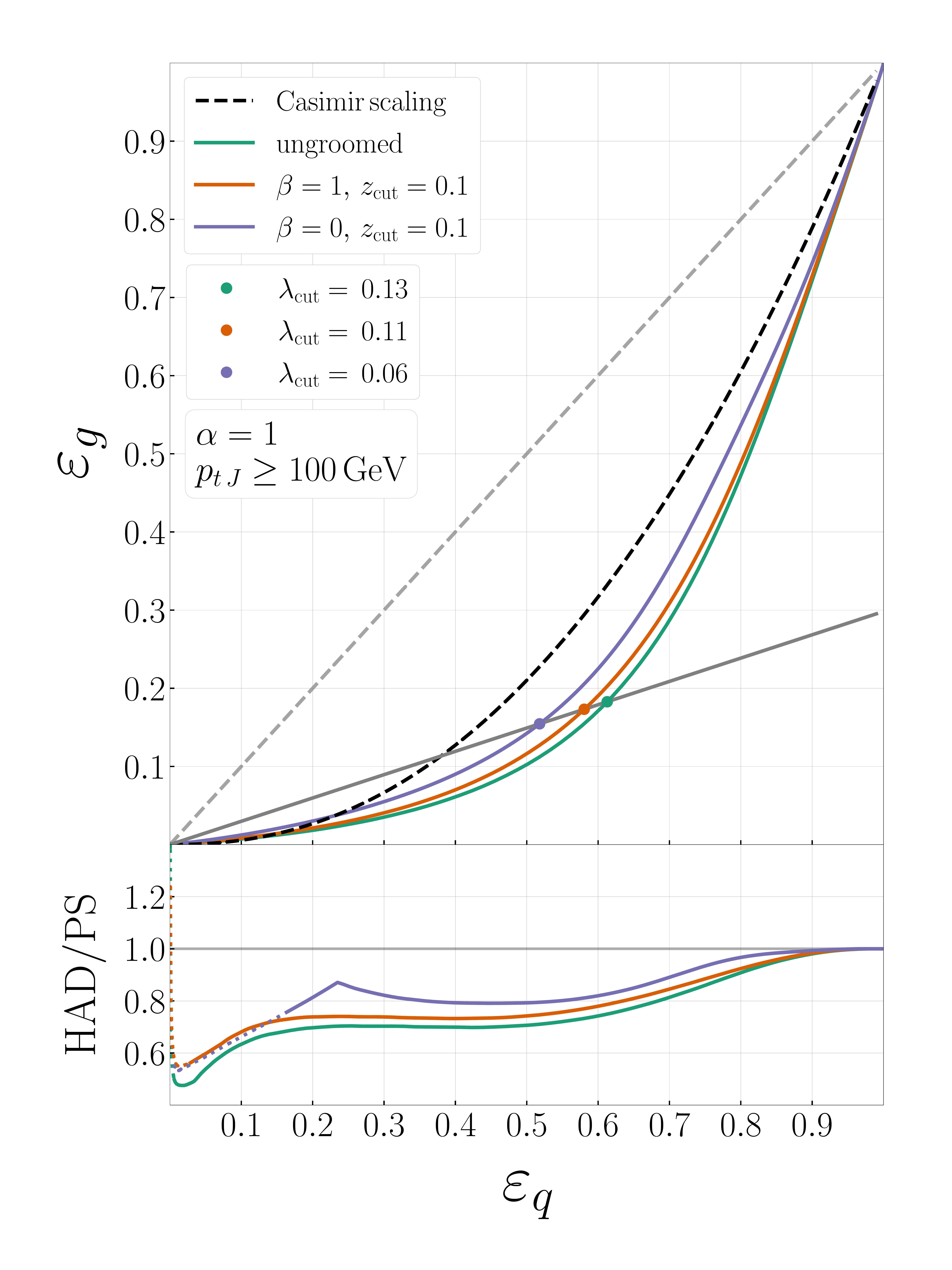}
  \captionof{figure}{The ROC curve for the angularity with $\alpha=1$ obtained with MC simulation using PYTHIA. Groomed and ungroomed cases are shown. From \cite{Caletti:2021ysv}.}
  \label{fig:roc}
\end{minipage}
\hspace{.05\linewidth}
\begin{minipage}{.45\linewidth}
  \includegraphics[width=\linewidth]{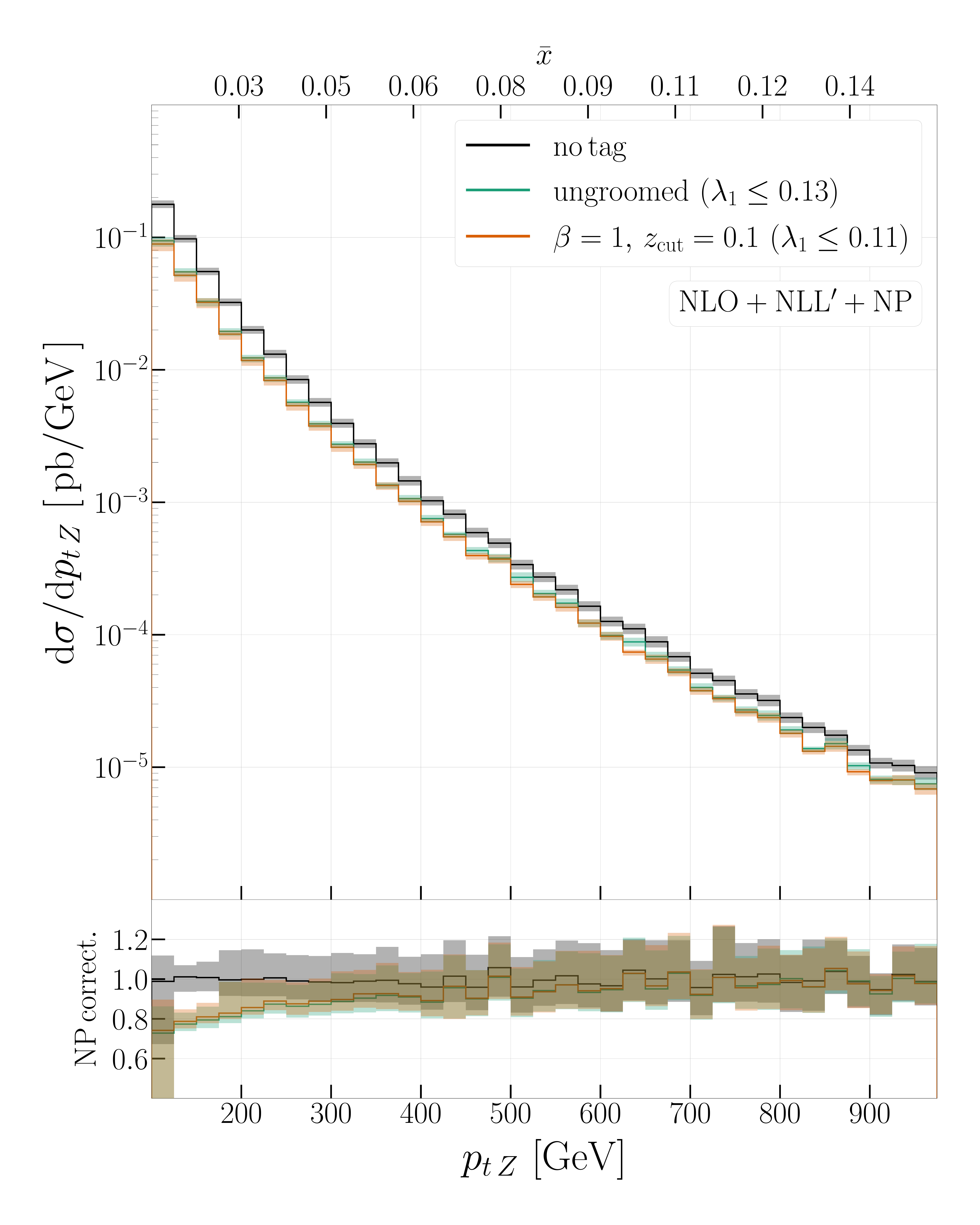}
  \captionof{figure}{The transverse momentum distribution of the Z boson in Z+jet events with the leading jet tagged as quark-initiated. The NLO+NLL' calculation is supplemented with a NP correction factor. From \cite{Caletti:2021ysv}.}
  \label{fig:pt}
\end{minipage}
\end{figure}

In order to determine the value of $\lambda_\text{cut}$ to perform such a cut on the jet angularity we plot the Receiver Operating Cheracteristic (ROC) curve by keeping the two partonic processes of interest separate. In particular
\begin{equation}
\varepsilon_k=\frac{1}{\sigma_{ij}}\int_0^{\lambda_{\text{cut}}}\frac{d\sigma_{ij}}{d\lambda}d\lambda\,, \hspace{1cm}\text{with}\hspace{1cm}i\,j\rightarrow k\,Z\,,
\end{equation}
for different values of $\lambda_{\text{cut}}$. To find the $\lambda_\text{cut}$ value we require a target value of the \emph{enhanced purity}, i.e. the initial-state gluon purity after a successfully applied tagging procedure (for quark-jet in the final state). This constraint is represented by the grey straight line in Fig. \ref{fig:roc}. The intersection of this constrain and a ROC curve determines a single point which corresponds to the selected value of $\lambda_{\text{cut}}$.

Now we have a good definition of quark/gluon jet accordingly to Section 1. Using such a tagger we can define the enhanced purity as 
\begin{equation}
\tilde{f_g}=\frac{\varepsilon_q\sigma_{qg}}{\varepsilon_g\sigma_{qq}+\varepsilon_q\sigma_{qg}}
\end{equation}
We plot the enhanced purity in Fig. \ref{fig:purity} for the groomed and ungroomed cases. It is clear we have a gain in purity which is roughly of a $10\%$.

In Fig. \ref{fig:pt} we provide our main result. Here we have the transverse momentum distributions in presence of tagging. Our calculation includes the resummation of logarithms of $\lambda_{\text{cut}}$ at NLL accuracy and is matched to NLO. 

\section*{Acknowledgement}
This work is supported by Università di Genova under the curiositydriven grant “Using jets to challenge the Standard Model of particle physics” and by the Italian Ministry of Research (MUR) under grant PRIN 20172LNEEZ.

\end{document}